# Dielectric properties of Poly(vinylidene fluoride) / $CaCu_3Ti_4O_{12}$ composites.

P.Thomas[a,b*], K.T.Varughese,[a] K.Dwarakanath,[a] and K.B.R.Varma.[b*]

[a] Dielectric Materials Division, Central Power Research Institute, Bangalore : 560 080, India

[b] Materials Research Centre, Indian Institute of Science, Bangalore: 560012, India,

**Abstract**

The possibility of obtaining relatively high dielectric constant polymer-ceramic composite by incorporating the giant dielectric constant material, $CaCu_3Ti_4O_{12}$ (CCTO) in a Poly(vinylidene fluoride) (PVDF) polymer matrix by melt mixing and hot pressing process was demonstrated. The structure, morphology and dielectric properties of the composites were characterized using X-ray diffraction, Thermal analysis, Scanning Electron Microscope, and Impedance analyzer. The effective dielectric constant ($\varepsilon_{eff}$) of the composite increased with increase in the volume fraction of CCTO at all the frequencies (100 Hz to 1MHz) under study. The dielectric loss did not show any variation upto 40 % loading of CCTO, but showed an increasing trend beyond 40%. The room temperature dielectric constant as high as 95 at 100Hz has been realized for the composite with 55 Vol % of CCTO, which has increased to about 190 at 150°C. Theoretical models like Maxwell's, Clausius-Mossotti, Effective medium theory, logarithmic law and Yamada were employed to rationalize the dielectric behaviour of the composite and discussed.

## 1. Introduction

Electronic systems in general consist of both the active and passive components. The technologies concerning the development of the passive components such as resistors, inductors, and capacitors are steadily growing in the electronic industries. Among these passive components, the capacitor is the one which attracts special attention due to its variety of functions that include decoupling, by-passing, filtering and timing capacitors. In recent years much work has been done on polymer-ceramic composites. Owing to the continuous development towards the miniaturization of electronics, high dielectric constant polymer-ceramic composites have become promising materials for embedded capacitor applications.

* **Corresponding author : Tel. +91-80-2293-2914; Fax: +91-80-2360-0683.**
  **E-mail : kbrvarma@mrc.iisc.ernet.in (K.B.R.Varma)**

Ferroelectric ceramics possessing very high dielectric constants are brittle and suffer from poor mechanical strength. On the other hand, polymers having low dielectric constants in the range of 2 to 5 traditionally used in low leakage capacitors are flexible, easy to process and possess high dielectric strength. New composites associated with high dielectric constant, and high dielectric breakdown strength to achieve high volume efficiency and energy storage density for applications of capacitors and electric energy storage devices [1-4] could be fabricated by combining the merits of polymers and ceramics. In order to achieve such an objective, high dielectric constant ferroelectric ceramics such as $Pb(Mg_{1/3}Nb_{2/3})O_3$-$PbTiO_3$(PMNPT), $Pb(Zr,Ti)O_3$(PZT) and $BaTiO_3$ (BT) have been used as fillers in polymers by previous workers [5-12]. Dielectric properties of 0-3 composites as a function of the volume fraction of the ceramic were studied by many authors [13-19]. Similarly, metallic fillers were also used to achieve ultra-high dielectric constant in the metal-polymer composites or three-phase composite systems [20,21]. Dielectric constants upto 200 were obtained in an irradiated PMNT-P(VDF-TrFE) 0-3 composite [7]. The most composite systems studied so far, were confined to ferroelectric ceramic fillers.

The $CaCu_3Ti_4O_{12}$ (CCTO) ceramic which has centrosymmetric *bcc* structure (space group Im3, lattice parameter a ≈ 7.391 $\overset{o}{A}$, and Z=2, has gained considerable attention due to its large dielectric constant (ε ~ $10^{4-5}$) which is nearly independent of frequency (upto 10 MHz) and low thermal coefficient of permittivity (TCK) over 100-600K temperature range [23,24]. Several schools of thoughts exist to explain the origin of high dielectric constant observed in CCTO ceramics [23,25-32]. Though several explanations have been put forward, the actual mechanism of the origin of giant dielectric constants in CCTO is still debatable as to whether it is intrinsic or extrinsic in nature. Recently, $CaCu_3Ti_4O_{12}$ (CCTO) powders have been used as a filler and studied to explore the possibility of obtaining high dielectric constant composites for potential capacitor applications [33-36]. It was reported that, the dielectric constant as high as 740 at 1kHz was achieved for a composition of fixed concentration : 50 vol % CCTO and 50 vol % PVDF-TrFE [33]. The dielectric constant increases as the CCTO content increases in the polymer and decreases as the frequency increases [34-36]. A three phase percolative composite with Aluminum powder yielded a high dielectric constant as high as 700 [34].

In this work we had used PVDF as matrix material because of its better thermal stability, tough and could be easily processed by injection mould/melt process. It has non-toxic, resistance to heat and chemicals and low water absorption characteristics which make it more suitable for making electronic components. The PVDF-CCTO composite fabricated for the first time by the melt mixing is reported in this work. In this article, the details pertaining to the fabrication and characterization of polymer-ceramic composites (PVDF/CCTO) involving CCTO as a filler and PVDF as matrix material are reported.

## 2. Experimental
### 2.1 Preparation of PVDF-CCTO composite.

The CCTO powders were prepared by conventional solid-state reaction route by heating a stoichiometric mixture of $CaCO_3$, $TiO_2$ and CuO at $1000^oC$ for 10h with intermediate repeated grinding [35]. The CCTO powders were ball milled for 10 h in a planetary mill using agate container to obtain micrometer (1-7 μm) sized particles. The Poly vinylidene fluoride (PVDF), molecular weight of MW 530,000, supplied by Sigma Aldrich was used in this work. For the fabrication of (PVDF/CCTO) composite, PVDF and CCTO powder were melt-mixed in a Brabender plasticorder (Model: PLE 331) for 15 min at $175^oC$, and then hot pressed at this temperature to obtain a sheet of 150 mm$^2$ with 0.5mm in thickness. A series of PVDF/CCTO composites with varying concentrations (0 to 55%) of CCTO by volume were fabricated.

## 2.2 Characterization Techniques

X-ray powder diffraction (XRD) studies were carried out using an XPERT-PRO Diffractometer (Philips, Netherlands) with Cu K$\alpha_1$ radiation ($\lambda = 0.154056$ nm) in a wide range of 2θ ($5^o \leq 2\theta \leq 85^o$). Thermo gravimetric (TG) analyses were done using the TA Instruments (UK), Model: TGA Q500, with alumina as the reference material. The experiments were carried out at a heating rate of $10^o$C min$^{-1}$ in flowing N$_2$ atmosphere (flow rate:50cm$^3$ min$^{-1}$). The microstructure and the morphology of the samples were characterized by using scanning electron microscope (SEM) (Cambridge Stereoscan S-360). For dielectric measurements, samples were polished using emery papers containing successively finer abrasives to achieve perfectly parallel and smooth surfaces. The surfaces of the specimens were electroded with silver paste and cured at $50^o$C. The capacitance of the electroded specimen was measured as a function of frequency (100Hz–1MHz) using an impedance gain-phase analyzer (HP4194A) at room temperature. A two terminal capacitor configuration was employed for this measurement. Theoretical models like Maxwell's, Clausius-Mossotti, Effective medium theory, logarithmic mixture rule and Yamada were employed to rationalize the dielectric behaviour of the composites under investigation.

## 3. Results and Discussion

Figure.1 (a-d) shows the X-ray powder diffraction patterns obtained for the starting powders (for synthesizing CCTO) calcined at 950$^o$C and 1000$^o$C, for different durations. Bragg peaks corresponding to the unreacted phases are noticed in the pattern even after 10 h of calcination at 950$^o$C. (Fig. 1(c)) Hence the processing temperature has been increased to 1000°C. Phase-pure CCTO could be obtained by calcining the powders at 1000°C/10h (Fig. 1(d)). The CCTO crystallites were found to be in the 1-7μm range. The X-ray diffraction pattern of the as received PVDF and the hot pressed composites (PVDF+55 vol %CCTO) are shown in Fig 2(a) and 2(b) respectively. It is evident from these patterns that the PVDF (Fig.2(a)) exists in mixed phases of $\alpha$ and $\gamma$ which are nonpolar. The sharp peaks at $2\theta$ corresponding to 17.7° (100), 18.4°(020), 19.9°(110) and 35.7°(200) are assigned to the $\alpha$-PVDF [6,10,11]. The broad peak at $2\theta = 26.8°(022)$ and 38.7°(211) could be indexed to the $\gamma$-phase of PVDF [37]. The X-ray diffraction pattern obtained for PVDF+55 vol %CCTO composite (Fig.2(b)), shows the predominance of CCTO phase. The intensities of the peaks pertaining to the PVDF are not significant. This is due to the presence of higher weight percent of CCTO than that of PVDF in the PVDF+55 vol %CCTO composite.

The thermo gravimetric analyses (TGA) were carried out to study the thermal stability of the polymer and the composites. Fig.3 shows the thermograms recorded for the pure PVDF and the composites. It is observed that the pure PVDF is stable up to 390°C and complete degradation of the polymer occurred at around 500°C. But the composites exhibited better stability than that of the pure polymer. As seen in the thermogram, complete degradation of the composites does not taken place as observed in the case of pure PVDF. The composites exhibit (Fig.3(b-e)) unique decomposition behaviour which is attributed to the presence of CCTO in PVDF. The weight loss obtained from the burnt (950°C/24h) out test on a few composite samples by the gravimetric determination is in good agreement with the experimental values.

The SEM micrographs of the hot-pressed as prepared composites containing various volume % of CCTO filler are shown in Fig.4(a-c). It is seen that the CCTO particle distribution is homogeneous with small trails of agglomeration associated with a smooth surface finish with minimum porosity. These results indicate that the CCTO is evenly distributed in the PVDF matrix. As reported earlier [38], good dispersion along with homogeneous packing of ceramic filler is likely to exhibit high dielectric constant. Indeed, we could achieve a dielectric constant as high as 95 at 100 Hz eventhough the distribution of the ceramic fillers are heterogeneous for the PVDF with 55 vol %CCTO composite.

The processing of such composites is much easier and could be made into a variety of shapes. The properties of such phase systems are governed by the number of factors: volume fraction of the dispersing phases, their properties and the way in which these different phases are interconnected [1]. The PVDF/ceramic composites are expected to show complex dielectric behaviour due to the different polymorphisms of PVDF.

Fig.5 presents the variation of dielectric constant and loss with frequency for the CCTO ceramic sintered at 1100/2h. The dielectric constant decreases as the frequency increases and at 10 kHz, the dielectric constant and the loss is 2500 and 0.04 respectively. The frequency dependence of the dielectric constant of the composite with different CCTO contents are shown in Fig. 6. As expected, the effective dielectric constant ($\varepsilon_{eff}$) increases with increase in CCTO content in PVDF at all the frequencies under study. In all the cases, the effective dielectric constants obtained are higher than that of pure PVDF, but much lower than that of pure CCTO. The low frequency dispersion increases with increase in the CCTO content. The reason for increased low frequency dispersion may be due to the space charge effects and high dielectric loss associated with CCTO. It is well known that PVDF has five different polymorphisms; of which four phases designated as $\alpha$, $\beta$, $\gamma$ and $\delta$ are stable at room temperature. The $\alpha$-phase is the most commonly present. The low dielectric constant as compared to that of CCTO observed for the PVDF/CCTO composite may be due to the following facts apart from the connectivity and particle size effects. Based on X-ray studies, it is confirmed that the PVDF is present in the mixed phases of $\alpha$ and $\gamma$ which are non-polar. Due to the non-polar nature of PVDF, and the constrained polymer chain hindering the contribution of electrical polarization, the value of $\varepsilon_r$ is lower than that of CCTO.

The frequency dependence of the dielectric loss of PVDF/CCTO composite with different volume % of CCTO are shown in the Fig.7. The loss undergoes two relaxations; one in the low frequency region and the other at high frequencies. The relaxation beyond 1MHz is related to the glass transition relaxation of PVDF and is denoted as $\alpha_a$ relaxation [39-41]. Similarly, the relaxation peak that appears below 100Hz could be attributed to $\alpha_a$ relaxation associated with molecular motion in the crystalline regions of PVDF [41]. It is observed that (Fig.7) the dielectric loss decreases in the 100 Hz to ~ 10 kHz frequency range and then subsequently increases up to a frequency (1MHz) that is covered in the present study. There is no appreciable change in the loss behavior upto the 40% of ceramic loading and then the loss increases especially at low frequencies when the ceramic loading is increased to 55% in the composite which is ascribed to the presence of CCTO particles in PVDF. It is believed that upto 40 % of ceramic loading, the uniform distribution of CCTO particles (as revealed by the scanning electron microscopy) in the immobile matrix, hinders the formation of networks and results in a decrease in the dielectric loss in the composite. However, at higher contents of CCTO, as seen in Fig. 4(c) there is a well established connectivity which resulted in an increase in the dielectric loss.

Figs 8 & 9 illustrate the frequency dependence of the dielectric constant and the dielectric loss of PVDF+55 vol % CCTO composite at different temperatures (50-150$^o$C) respectively. The dielectric constant increases with increasing temperature (Fig.8). The room temperature dielectric constant is 95 at 100Hz, which has increased to about 190 at 150$^o$C (100Hz). But the rise in dielectric constant with rise in temperature decreases with increase in frequency as shown in the Fig.8. There is no significant change in the dielectric constant in the 100kHz to 1MHz frequency range even at 150$^o$C. These results are consistent with the fact that the dielectric constant of CCTO is nearly independent of frequency (upto 1MHz) and has low thermal coefficient of permittivity (TCK) over a wider range of temperatures (100-600K) [27,28]. The dielectric loss (Fig.9) also increases as the temperature increases

and decreases as the frequency increases which is also consistent with that of the CCTO behaviour.

In the present work, various models are used in order to predict the effective dielectric constant of the composite. Fig.10 shows the room temperature dielectric constant of the composite at 10 kHz (experimental) for different volume fractions of CCTO. For comparison, the dielectric constant calculated based on various models [42-46] are also included in the same figure. The dielectric property of a diphasic dielectric mixture comprising of spherical crystallites with high dielectric constant dispersed in a matrix of low dielectric constant could be well described by Maxwell's model [42]. According to this model the effective dielectric constant of the composite is given by

$$\varepsilon_{eff} = \left( \frac{\delta_p \varepsilon_p \left( 2/3 + \varepsilon_c/3\varepsilon_p \right) + \delta_c \varepsilon_c}{\delta_p \left( 2/3 + \varepsilon_c/3\varepsilon_p \right) + \delta_c} \right) \qquad (1)$$

where, $\varepsilon_c$, $\varepsilon_p$, $\delta_c$ and $\delta_p$ are the dielectric constants of CCTO, PVDF, the volume fraction of the dispersoid and the polymer, respectively. The values that are substituted for $\varepsilon_c$ and $\varepsilon_p$ are 2500 and 6.4 (at 10kHz). Here, the predicted value of $\varepsilon_{eff}$ deviates much from that of the experimental value for all the volume fractions of CCTO under study.

The most commonly used dielectric mixture rule is Lichtenecker's [43] which is also referred to as the logarithmic mixture rule.

$$\log \varepsilon_{eff} = \delta_1 \log \varepsilon_1 + \delta_2 \log \varepsilon_2 \qquad (2)$$

This mixture rule is the intermediate form of the series and parallel combination laws for dielectric mixture. In the lower volume fraction regime, the experimental results are comparable with those obtained using this model. The $\varepsilon_{eff}$ value for $f_{CCTO}$=0.2, 0.3 is 21.1 and 38.3, which are comparable to the experimental $\varepsilon_{eff}$ (24.7 and 42.2). At higher volume %, predicted value of $\varepsilon_{eff}$ deviates much from that of the experimental value (Fig.10).

Models like Clausius-Mossotti [44] and Effective Medium theory [45] were also used for predicting the effective dielectric constant. In the Clausius-Mossotti treatment of a mixture of dielectrics composed of spherical crystallite dispersed in a continuous medium, the effective dielectric constant ($\varepsilon_{eff}$) of the composite is calculated using the equation,

$$\varepsilon_{eff} = \varepsilon_p \left[ 1 + 3\delta_c \left( \frac{(\varepsilon_c - \varepsilon_p)}{\varepsilon_c + 2\varepsilon_p} \right) \right] \quad (3)$$

Since CCTO particles that are dispersed in the matrix are non-spherical in nature as evidenced by SEM, the predicted value of $\varepsilon_{eff}$ does not fit in to the model and is at variance with that of the experimental value (Fig.10).

The effective medium theory (EMT) model [45] has been established taking into account the morphology of the particles. According to which the effective dielectric constant is given by

$$\varepsilon_{eff} = \varepsilon_p \left[ 1 + \frac{f_c(\varepsilon_c - \varepsilon_p)}{\varepsilon_p + n(1 - f_c)(\varepsilon_c - \varepsilon_p)} \right] \quad (4)$$

where $f_c$, is the volume fraction of the ceramic dispersed, $\varepsilon_c$, $\varepsilon_p$ and $n$ are the dielectric constants of the particle, polymer and the ceramic morphology fitting factor respectively. The small value of $n$ indicates the filler particles to be in near-spherical shape, while a high value of $n$ indicates largely non-spherically shaped particles. In the low volume fraction regime, the $\varepsilon_{eff}$ value for $f_{CCTO}$=0.2 is 20.4, which is comparable to the experimental $\varepsilon_{eff}$ (24.7). The ceramic particle size has been proven to influence the effective dielectric constant in the composite and the size of the CCTO particle used in this work is around 1-7μm range. Though this model is suitable for the ceramic particles less than 1μm and also depends on the their morphology, interestingly, the experimental value fits well into the EMT model [45] with the shape parameter $n$=0.11. The difference between the experimental data and the predicted value is less than 10 % in the higher volume fraction regime (upto 55 vol%) (Fig.10). It is to be noted that the morphology fitting factor obtained from the fit ($n$=0.11) is closer to that of reported ($n$=0.13) [45]. The theoretical morphology fitting factor is closer to that of the experimental one, and is consistent with that observed (irregular shaped particles) in the present studies (Fig4(a-c)) [45]. Therefore, the close agreement that is found between the experimental and theoretical values is attributed to the morphology of the dispersed particles.

The model that was developed by Yamada [46], is used to predict $\varepsilon_{eff}$. According to which,

$$\varepsilon_{eff} = \varepsilon_1 \left[ 1 + \frac{n f_{CCTO}(\varepsilon_2 - \varepsilon_1)}{n\varepsilon_1 + (\varepsilon_2 - \varepsilon_1)(1 - f_{CCTO})} \right] \quad (5)$$

Here, $\varepsilon_1$ and $\varepsilon_2$ are, respectively, the dielectric constant of the epoxy and CCTO ceramics, "$n$" is the parameter related to the geometry of the ceramic particles and $f_{CCTO}$ is the volume fraction of the CCTO phase in the matrix. The parameter "$n$" was evaluated to fit the theoretical value obtained from the eqn (5) to the observed values. Here, in the higher volume fraction regime, the experimental results are comparable to those obtained using this model, when the parameter "$n$" is around 9.3 (Fig.10). The $\varepsilon_{eff}$ value for $f_{CCTO}$=0.4 is 43.7 which is comparable to the experimental $\varepsilon_{eff}$ (48.2). Similarly, the $\varepsilon_{eff}$ value for $f_{CCTO}$=0.55 is 74 which is comparable to the experimental $\varepsilon_{eff}$ (69.4). Here again, it is observed that the "$n$" parameter is related to the geometry of the particles, and the "$n$" parameter of 9.3 obtained from the fit is correlated to the irregular geometry of the ceramic particles as mentioned above. The particles geometry in turn effect the connectivity and hence the observed effective dielectric constant in the composite. These observations clearly demonstrate the effective applicability of this model to rationalize the dielectric characteristics of the present composite at higher filler content.

Mostly, in the polymer-ceramic composite systems, the ceramic is in the powder form instead of sintered form. The dielectric constant of the ceramic in the sintered form is much higher than that of the un sintered one. The PVDF-CCTO composite in this work deals with the a composite of medium composed of dispersed unsintered CCTO powder within the polymer. Hence, at higher volume fraction, the experimental results are comparable to those obtained using the EMT and Yamada models.

The effective dielectric constant of polymer /filler composite material is dependent not only on the dielectric constants of the polymer and the filler, size and shape of the filler and the volume fraction of the filler, but also on the dielectric constant of the interphase region, volume of the interphase region and on the type of coupling agents [47-50]. One cannot afford to ignore these aspects for fabricating composites with improved dielectric properties.

## 4. Conclusions

Composites of CCTO /PVDF with CCTO content ranging from 0% to 55% by volume were fabricated and characterized. The flexibility of the composite could be retained when the ceramic loading was upto 30 vol %. As expected, the effective dielectric constant ($\varepsilon_{eff}$) and dielectric loss increases with increase in the volume fraction of CCTO at all frequencies under study. Dielectric constant as high as 95 at 100Hz was achieved. Though the dielectric constant increases with increasing temperature, the rise in dielectric constant with rise in temperature is not significant at frequencies from 10kHz to 1MHz. The SEM of the composites revealed the excellent distribution of CCTO fillers in PVDF matrix.

Among the various models used for rationalizing the dielectric behaviour, experimental $\varepsilon_{eff}$ at higher volume fractions regime is comparable with those obtained with the models like EMT ( $n$=0.11) and Yamada ($n$ =9.3) and the difference between the experimental data and the predicted value is less than 10 % .

However, all these models have the limitations that interfacial structure and chemistry in a composite have not been taken into account. Since the microstructure and microchemistry of interfaces often controls the mechanical, physical and electrical properties of composite materials, incorporating interfacial effect, the dielectric behaviour of CCTO/PVDF composite needs to be studied in detail. The CCTO/PVDF composite in this study have further scope of fabricating modified composites with treatment of coupling agent with CCTO, for its interfacial structural properties.

**Acknowledgements**

The management of Central Power Research Institute is acknowledged for the financial support (CPRI Project No.5.4.49).

**FIGURE CAPTIONS**

Figure.1. X-ray diffraction patterns for the conventionally synthesized powder at (a) $950^{o}C/60min$ (b) $950^{o}C/6h$, (c) $950^{o}C/10h$ and (d) $1000^{o}C/10h$ –Phase-pure CCTO.

Figure.2. X-ray diffraction pattern for the hot pressed (a) PVDF and (b) PVDF+55 Vol % CCTO

Figure.3 Thermal analysis (TG) for the (a) pure PVDF, (b) 30 (c) 40 (d) 50 and (e) for 55 Vol % CCTO-PVDF composite.

Figure.4. Scanning electron micrographs of CCTO/PVDF composite for different volume percent of CCTO (a) 40, (b) 50, and (c) 55 % .

Figure.5. Frequency dependence of dielectric constant and D.loss (measured at 300K) for the CCTO ceramic sintered at 1100/2h.

Figure.6. Frequency dependence of effective dielectric constant ($\varepsilon_{eff}$) (measured at 300K) of CCTO/PVDF composite as a function of volume percent of CCTO ($f_{CCTO}$).

Figure.7. Frequency dependence of effective dielectric loss (D) (measured at 300K) of CCTO/PVDF composite as functions of volume percent of CCTO ($f_{CCTO}$)

Figure. 8. Temperature dependant of dielectric constant for PVDF+55 Vol % CCTO composite at a given frequencies.

Figure. 9. Temperature dependant of dielectric loss for PVDF+55 Vol % CCTO composite at a given frequencies.

Figure.10. Variation of effective dielectric constant ($\varepsilon_{eff}$) (measured at 300K and 10kHz of CCTO/PVDF composite as a function of volume percent of CCTO particles ($f_{CCTO}$). For comparison, the calculations by using various models are also shown.

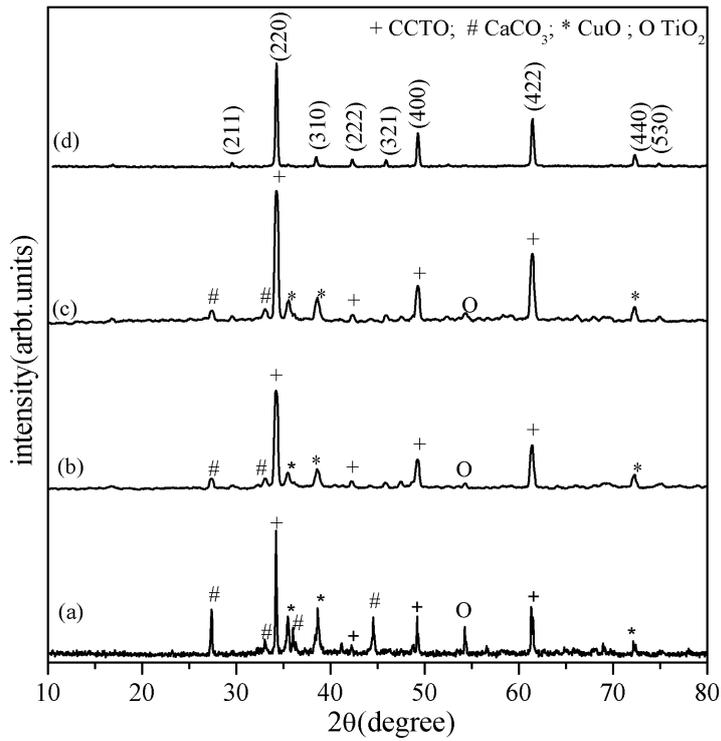

Figure.1. X-ray diffraction patterns for the conventionally synthesized powder at (a) 950°C/60min  (b) 950°C /6h, (c) 950°C /10h  and (d) 1000°C/10h –Phase-pure CCTO.

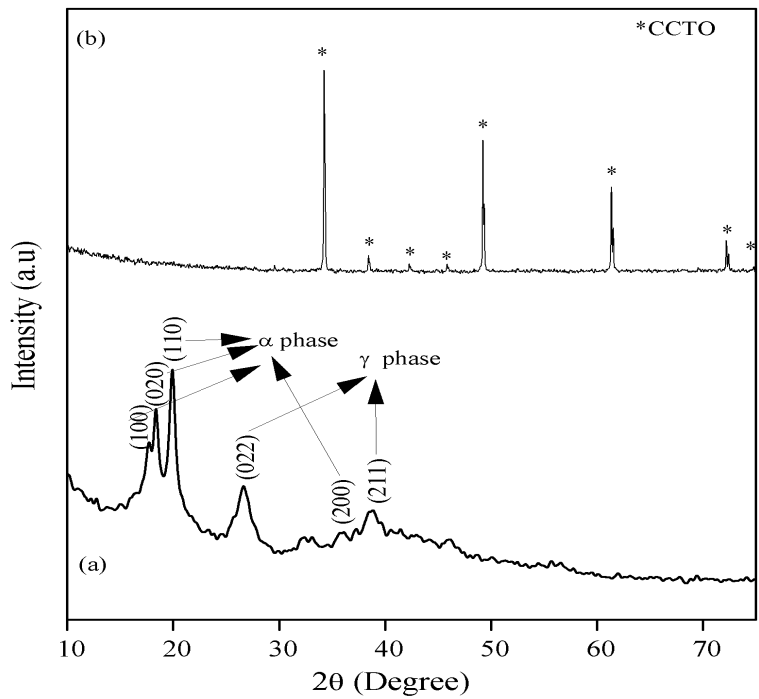

Figure.2. X-ray diffraction pattern for the hot pressed (a) PVDF and (b) PVDF+55 Vol % CCTO

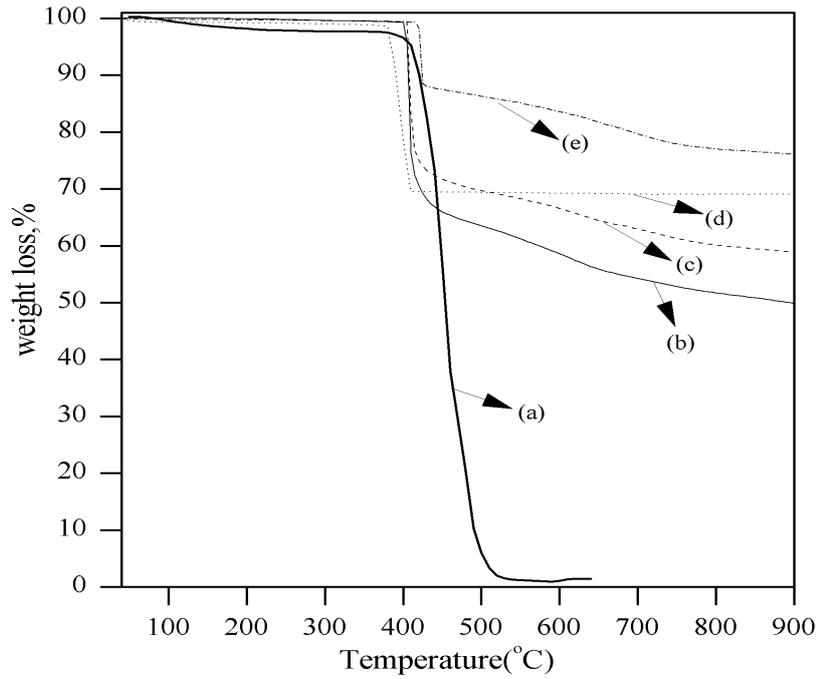

Figure.3 Thermal analysis (TG ) for the (a) pure PVDF, (b) 30 (c) 40 (d) 50 and (e) for 55 Vol % CCTO-PVDF

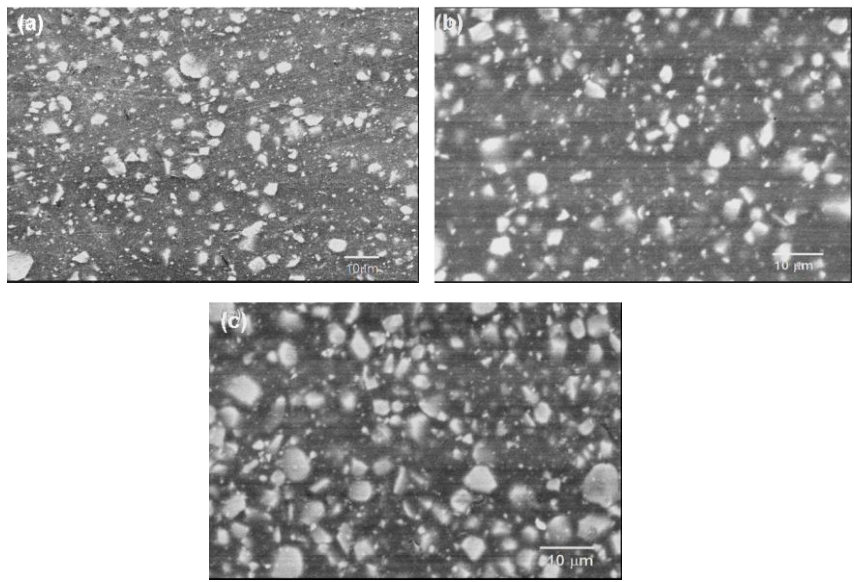

Figure.4. Scanning electron micrographs of CCTO/PVDF composite for different volume percent of CCTO (a) 40, (b) 50, and (c) 55 vol% .

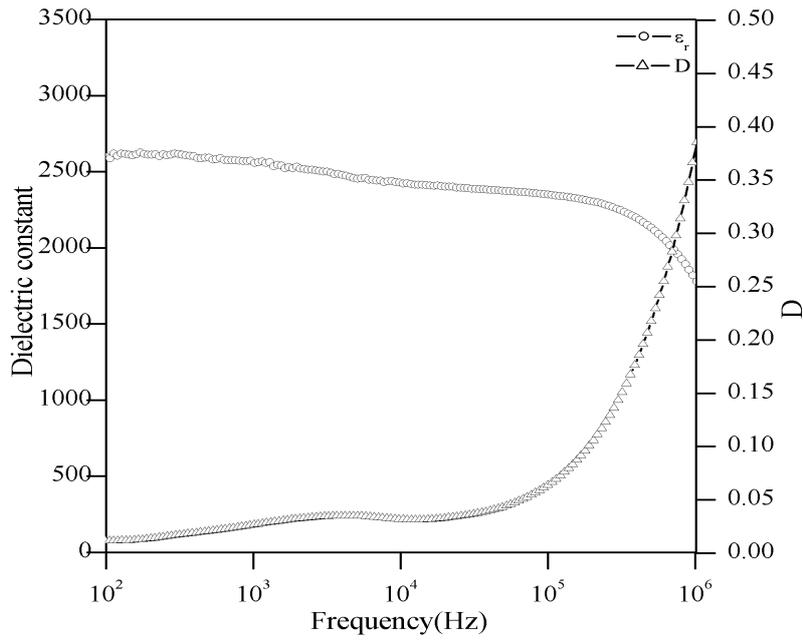

Figure.5. Frequency dependence of dielectric constant and D.loss (measured at 300K) for the CCTO ceramic sintered at 1100/2h.

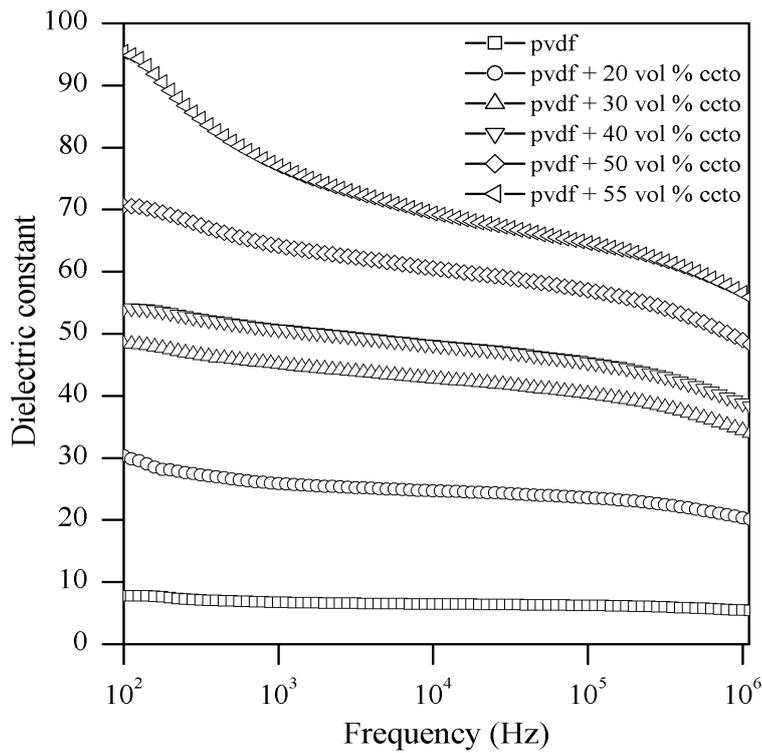

Figure.6. Frequency dependence of effective dielectric constant ($\varepsilon_{eff}$) (measured at 300K) of CCTO/PVDF composite as a function of volume percent of CCTO ($f_{CCTO}$).

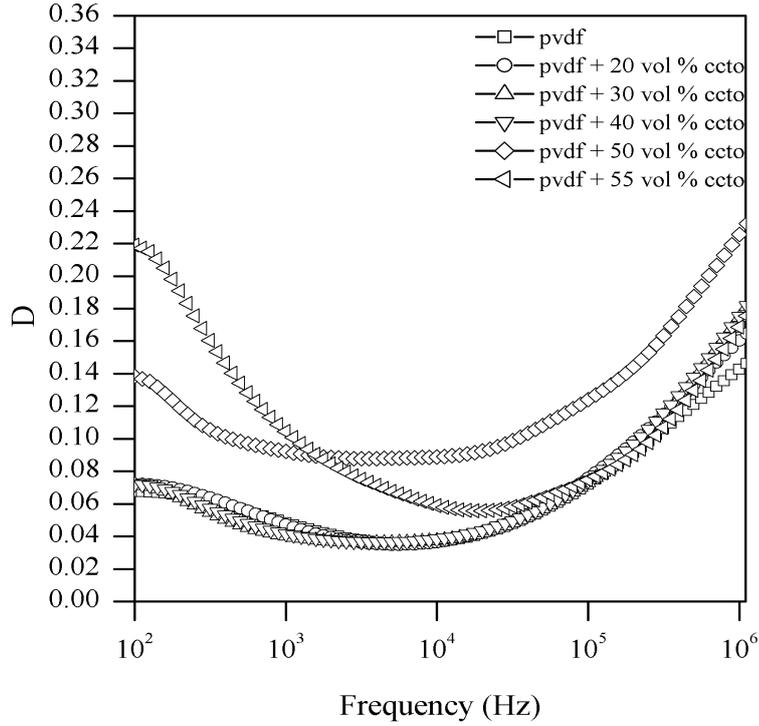

Figure.7. Frequency dependence of effective dielectric loss (D) (measured at 300K) of CCTO/PVDF composite as functions of volume percent of CCTO ($f_{CCTO}$)

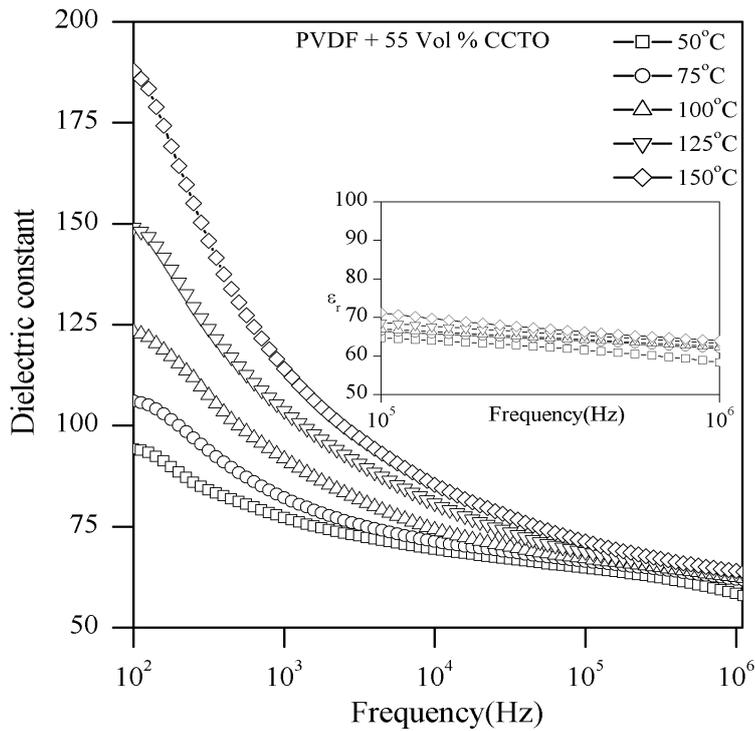

Figure. 8. Temperature dependant of dielectric constant for PVDF+55 Vol % CCTO composite at a given frequencies.

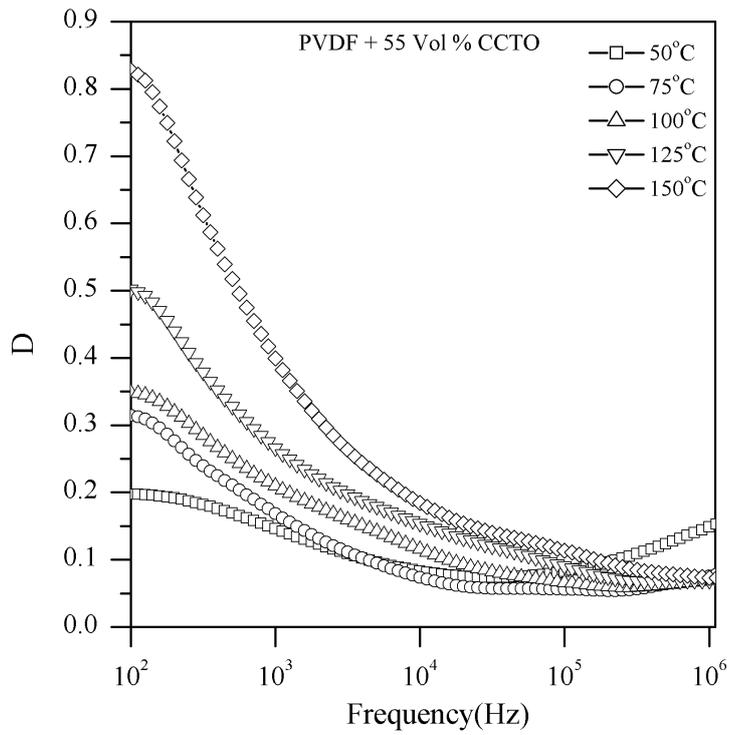

Figure. 9. Temperature dependant of dielectric loss for PVDF+55 Vol % CCTO composite at a given frequencies.

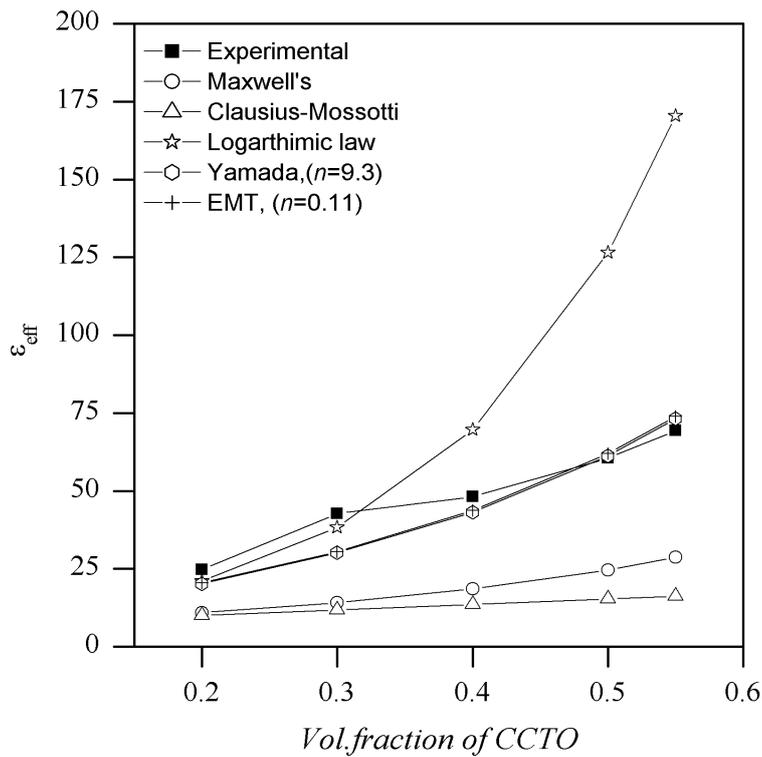

Figure.10. Variation of effective dielectric constant ($\varepsilon_{eff}$) (measured at 300K and 10kHz of CCTO/PVDF composite as a function of volume percent of CCTO particles ( $f_{CCTO}$). For comparison, the calculations by using various models are also shown.